\documentclass[aps,pra,twocolumn,floats,amsmath,amssymb,superscriptaddress]{revtex4-1}
\usepackage[utf8]{inputenc}
\usepackage{graphicx}
\usepackage{epsfig}
\usepackage{amsfonts}
\usepackage{amsmath}
\usepackage{natbib}
\usepackage{multirow}
\usepackage{bm}
\usepackage{hyperref}
\hypersetup{
    colorlinks=true,
    linkcolor=green,
    filecolor=green,
    citecolor=green,
    urlcolor=cyan,
}
\usepackage{epstopdf}
\usepackage{ulem}
\usepackage{xcolor}  

\definecolor{turq}{rgb}{0.0, 0.72, 0.92}

\begin{document}
\title{Fingerprints of optical absorption in the perovskite LaInO$_{3}$: Insight from many-body theory and experiment}
\author{Wahib Aggoune}
\affiliation{Institut f\"{u}r Physik and IRIS Adlershof, Humboldt-Universit\"{a}t zu Berlin, 12489 Berlin, Germany}
\author{Klaus Irmscher}
\affiliation{Leibniz-Institut f\"{u}r Kristallz\"{u}chtung, Max-Born-Str. 2, 12489 Berlin, Germany}
\author{Dmitrii Nabok}
\author{Cecilia Vona}
\affiliation{Institut f\"{u}r Physik and IRIS Adlershof, Humboldt-Universit\"{a}t zu Berlin, 12489 Berlin, Germany}
\affiliation{European Theoretical Spectroscopic Facility (ETSF)}
\author{Saud Bin Anooz}
\affiliation{Leibniz-Institut f\"{u}r Kristallz\"{u}chtung, Max-Born-Str. 2, 12489 Berlin, Germany}
\author{Zbigniew Galazka}
\affiliation{Leibniz-Institut f\"{u}r Kristallz\"{u}chtung, Max-Born-Str. 2, 12489 Berlin, Germany}
\author{Martin Albrecht}
\affiliation{Leibniz-Institut f\"{u}r Kristallz\"{u}chtung, Max-Born-Str. 2, 12489 Berlin, Germany}
\author{Claudia Draxl}
\affiliation{Institut f\"{u}r Physik and IRIS Adlershof, Humboldt-Universit\"{a}t zu Berlin, 12489 Berlin, Germany}
\affiliation{European Theoretical Spectroscopic Facility (ETSF)}
\date{\today}
 
\begin{abstract}
 We provide a combined theoretical and experimental study of the electronic structure and the optical absorption edge of the orthorhombic perovskite LaInO$_{3}$. Employing density-functional theory and many-body perturbation theory, we predict a direct electronic quasiparticle band gap of about 5 eV and an effective electron (hole) mass of 0.31 (0.48) m$_{\mathrm{0}}$. We find the lowest-energy excitation at 0.2 eV below the fundamental gap, reflecting a sizeable electron-hole attraction. Since the transition from the valence band maximum (VBM, $\Gamma$ point) is, however, dipole forbidden the onset is characterized by weak excitations from transitions around it. The first intense excitation appears about 0.32~eV above. Interestingly, this value coincides with an experimental value obtained by ellipsometry (4.80 eV) which is higher than the onset from optical absorption spectroscopy (4.35 eV). The latter discrepancy is attributed to the fact that the weak transitions that define the optical gap are not resolved by the ellipsometry measurement. The absorption edge shows a strong dependency on the light polarization, reflecting the character of the involved valence states. Temperature-dependent measurements show a redshift of the optical gap by about 120 meV by increasing the temperature from 5 to 300 K. Renormalization due to zero-point vibrations is extrapolated from the latter measurement to amount to 150 meV. By adding the excitonic binding energy of 0.2 eV obtained theoretically to the experimental optical absorption onset, we determine the fundamental band gap at room temperature to be 4.55 eV. 
\end{abstract}
\maketitle


\section{Introduction}

Perovskite oxides have emerged to play a key role in the next generation of technological devices, owing to their versatility, spanning insulators, semiconductors, as well as conductors, and exhibiting other characteristics such as superconductivity, ferromagnetism, or ferroelectricity. The perovskite oxide structure has the general formula ABO$_{3}$ (A and B are cations; O is the anion oxygen). Among them, perovskites with a large optical gap accompanied with high electrical mobility are found, making them ideal transparent conducting oxides (TCOs). Such characteristics are in high demand for various electronic applications. The cubic perovskite BaSnO$_{3}$ has appeared as the most promising TCO due to its high room-temperature mobility, reaching 320 cm$^{2}$(Vs)$^{-1}$ when doped with lanthanum, and its high transparency~\cite{Hkim+12ape}. Recently, polar-discontinuity doping has been suggested~\cite{Useong+15apl} as an effective way to benefit from BaSnO$_{3}$'s characteristics, which leads to the formation of a high-mobility two-dimensional electron gas (2DEG). This can be achieved in a heterostructure combining the non-polar BaSnO$_{3}$ with polar oxide perovskites. Among them, orthorhombic LaInO$_{3}$ has emerged as the most promising candidate for such heterojunctions, due to the relatively small lattice mismatch in its pseudo-cubic structure ($<$0.02$\%$) and a favorable conduction-band offset to confine the 2DEG within the BaSnO$_{3}$ side~\cite{Useong+15apl,kim+18apl,kim+19sp,claudialau+19apl,Martina+20prm}. As such, LaInO$_{3}$ has been explored as a gate oxide in a field-effect transistor based on a doped BaSnO$_{3}$ channel~\cite{Useong+15apl}. From an experimental point of view, LaInO$_{3}$ is considered as the best substrate to overcome difficulties related to the growth of the BaSnO$_{3}$ crystals~\cite{jang+17jap}.

In contrast to BaSnO$_{3}$ that has been intensively investigated from both theory and experiment~\cite{mizoguchi+04jacs,Hkim+12prb,galazka+16jpcm,beom+17cap,dongmin+14apl,chambers+16apl,Krish+16apl,krish+17prb}, there are only a few studies devoted to LaInO$_{3}$, reporting a large direct fundamental band gap, however, with contradicting results of about 5 eV \cite{Useong+15apl} and 4.13 eV \cite{jang+17jap}. Available density-functional-theory (DFT) calculations based on the generalized gradient approximation (GGA) predict -- as expected -- strongly underestimated band gaps of 2.55 eV ~\cite{aytac+16pm} (indirect gap) and 3.1 eV~\cite{Useong+15apl}. Computed effective electron and hole masses~\cite{Useong+15apl} are about 0.41 and 0.53 m$_{\mathrm{0}}$, respectively. Importantly, neither origin nor character of the absorption onset are clear. Weak absorption below the main onset~\cite{jang+17jap} suggested that such transitions could be an intrinsic property of orthorhombic perovskites with fully occupied \textit{d} orbitals. More recently~\cite{Zbigniew+20jcg}, an optical gap of about 4.35 eV was reported, where the observed weak transitions below the main onset were attributed to point-defects.

In this work, we provide the characteristics of ideal crystalline LaInO$_{3}$ in terms of its electronic and optical properties, employing a state-of-the-art \textit{ab initio} approach based on many-body perturbation theory (MBPT). The $G_0W_0$~\cite{hedi65pr,hybe-loui85prl} approximation is used to compute the quasiparticle (QP) band structure based on semilocal DFT as well the hybrid functional HSE06~\cite{HSE+06jcp}. We obtain the optical absorption spectrum using the Bethe-Salpeter equation (BSE)~\cite{hank-sham80prb,stri88rnc,rohl-loui00prb,pusc-ambr02prb}) that allows us to assess the role of electron-hole (e-h) interaction. We unravel the origin of the absorption edge, resolving discrepancies regarding the optical gap determined by either optical absorption or ellipsometry as well as the dependence of the absorption onset on the light polarization. Moreover, we determine the band-gap renormalization by zero-point vibrations (ZPV) and analyze the characteristics of the low-lying excitons in this material. 


\section{Theoretical and experimental details}
\subsection{Theory}
\label{method}
Ground-state (GS) properties are calculated using DFT~\cite{hohe-kohn64pr,kohn-sham65pr} with the GGA approximation in the PBEsol parametrization~\cite{PBEsol+08prl} of the xc functional. The hybrid functional HSE06~\cite{HSE+06jcp} with 25\% of exact (Hartree-Fock) exchange is also employed for comparison. 
 
QP energies are computed within the $G_{0}W_{0}$ approximation \cite{hedi65pr,hybe-loui85prl} as 
\begin{equation}
\varepsilon_{i}^{QP}=\varepsilon_{i}
^{KS}+\langle\phi_{i}^{KS}\vert\Sigma(\varepsilon_{i}^{QP})
-v_{xc}^{KS}\vert\phi_{i}^{KS}\rangle,
\end{equation}
where $\Sigma$ is the non-local and energy dependent electronic self-energy, $\varepsilon_{i}^{KS}$ and $\phi_{i}^{KS}$ are the Kohn-Sham energies and wave-functions, respectively, and $v_{xc}^{KS}$ represents the xc potential. Band structure and effective masses are computed by making use of Wannier interpolation~\cite{seb+20prb}.

The optical spectra are obtained by solving the BSE, the equation of motion of the two-particle Green function~\cite{hank-sham80prb,stri88rnc,rohl-loui00prb}. This problem can be mapped onto the secular equation 
\begin{equation}
\sum_{v'c'\mathbf{k'}} H^{BSE}_{vc\mathbf{k},v'c'\mathbf{k'}}A^{\lambda}_{v'c'\mathbf{k'}} = E^{\lambda}A^{\lambda}_{vc\mathbf{k},}
\label{eq:ham}
\end{equation}
where $v$, $c$, and \textbf{k} indicate valence bands, conduction bands, and \textbf{k}-points in reciprocal space, respectively. The effective Hamiltonian consists of three terms, $H^{BSE} = H^{diag} + H^{dir} + 2H^{x}$. The first term, $H^{diag}$, accounts for \textit{vertical} transitions between QP energies and, when considered alone, corresponds to the independent approximation (IQPA). The other two terms incorporate the screened Coulomb interaction ($H^{dir}$) and the bare electron-hole exchange ($H^{x}$). The factor 2 in front of the latter accounts for the spin multiplicity in non-spin-polarized systems. The eigenvalues of Eq.~\eqref{eq:ham}, $E^{\lambda}$, are the excitation energies. The corresponding eigenvectors, $A^{\lambda}_{vc\mathbf{k}}$, provide information about the composition of the $\lambda$-th excitation and act as weighting factors in the transition coefficients
\begin{equation}
\mathbf{t}_{\lambda} = \sum_{vc\mathbf{k}} A^{\lambda}_{v c \mathbf{k}} \frac{\langle v \mathbf{k} \vert \widehat{\mathbf{p}} \vert c \mathbf{k} \rangle}{\epsilon_{c \mathbf{k}}\ -\ \epsilon_{v \mathbf{k}}},
\label{eq:osci} 
\end{equation}
that determine the oscillator strength in the imaginary part of the macroscopic dielectric function,
\begin{equation}
\mathrm{Im}\varepsilon_M~=~\dfrac{8\pi^2}{\Omega} \sum_{\lambda} |\mathbf{t}_{\lambda}|^2 \delta(\omega - E^{\lambda}),
\label{eq:abs}
\end{equation}
where $\Omega$ is the unit cell volume. 
We introduce the weights characterizing to which extent valence and conduction states at a given $\mathbf{k}$-point contribute to a transition as:
  \begin{equation}
\omega_{v\mathbf{k}}^{\lambda}=\sum_{c}|A_{vc\mathbf{k}}^{\lambda}|^{2},~~\omega_{c\mathbf{k}}^{\lambda}=\sum_{v}|A_{vc\mathbf{k}}^{\lambda}|^{2}. 
\label{eq:weight}
  \end{equation}
All calculations are performed using \texttt{exciting}~\cite{gula+14jpcm,nabo+16prb,vor+es19}, an all-electron full-potential code, implementing the family of linearized augmented planewave plus local orbitals [(L)APW+LO] methods. For the atomic species involved, namely lanthanum (La), indium (In), and oxygen (O), muffin-tin radii (R$_{MT}$) of 2.2, 2.0, and 1.6 bohr, respectively, are adopted. The GS calculations are carried out with the LAPW method using a basis-set cutoff of R$_{MT}$G$_{max}$=8, where R$_{MT}$ here refers to the radius of the smallest sphere (1.6 bohr), {\it i.e.}, G$_{max}$=5. To obtain numerically precise electronic properties, local orbitals of three \textit{s}, three \textit{p}, and four \textit{d} shells for La; four \textit{s}, three \textit{p}, and four \textit{d} shells for In; and three \textit{s} and three \textit{p} shells for oxygen are added to the LAPW basis set. The In 4\textit{s} and 4\textit{p} orbitals are treated with additional LOs as semicore states. The sampling of the Brillouin zone (BZ) is carried out with a 6 $\times$ 6 $\times$ 4 $\textbf{k}$-grid. These parameters ensure a numerical precision of less than 10~meV in both the total energy and the PBEsol band gap. The atomic positions are relaxed until the residual forces on each atom are less than 0.005 eV/\AA{}. To account for spin-orbit coupling, we employ a second-variational procedure. The latter is performed using a slightly reduced basis-set cutoff of R$_{MT}$G$_{max}$=6.5 and including 100 empty states.

For the HSE06 calculations, a basis-set cutoff of R$_{MT}$G$_{max}$=6 is used. Employing 400 empty states and a 4 $\times$ 4 $\times$ 2 $\textbf{k}$-mesh, a numerical precision of about 50 meV is reached for the band gap. Using the same parameters, $G_0W_0$ calculations are performed on top of the HSE06 groundstate ($G_{0}W_{0}$@HSE06). The numerical uncertainty of the QP band gap is estimated to be less than 50 meV. Also for the solution of the BSE~\cite{vor+es19}, a plane-wave cutoff R$_{MT}$G$_{max}$=6 is adopted. 

The screened Coulomb potential is computed using 90 empty bands. In the construction of the BSE Hamiltonian 22 occupied and 18 unoccupied bands are included, and a shifted 8 $\times$ 8 $\times$ 6 $\textbf{k}$-point mesh is adopted. Since the excitations at the onset are built by a small group of top valence and bottom conduction states that are nearly identical in PBEsol and $G_{\mathrm{0}}W_{\mathrm{0}}@$HSE06, as we will see in Section~\ref{electronic}, these calculations are based on the PBEsol band structure, applying a scissor shift to reproduce the $G_{\mathrm{0}}W_{\mathrm{0}}@$HSE06 gap. Our choice of parameters ensures converged spectra up to 6 eV, and a numerical precision of less than 30~meV for the binding energy of the lowest-energy excitons. A Lorentzian broadening of 0.1 eV is applied to the spectra. Atomic structures and isosurfaces are visualized using the VESTA software \cite{momm-izum11jacr}. 

\subsection{Experiment}
Bulk LaInO$_{3}$ single crystals were grown by the vertical gradient freeze method. Details of the growth procedure were reported in Ref.~\cite{Zbigniew+20jcg}. Wafers of up to 10$\times$10 mm$^{2}$ size could be prepared from large single crystal grains. Stoichiometric composition, phase purity, and structural quality were substantiate in Ref.~\cite{Zbigniew+20jcg}. For the optical measurements two wafers of high structural perfection were selected, {\it i.e.}, their x-ray rocking curves had full width at half maximum well below 80 arcsec. The samples had lateral dimensions of about 5$\times$5 mm$^{2}$ and were thinned down to 80 $\mu$m thickness. The surface orientation of the wafers was (100) and (001), respectively, such that for perpendicularly incident, linearly polarized light the direction of the electric field vector \textbf{E} could be chosen parallel to either of the three principal crystal axes. For the optical absorbance measurements, the samples were double-side polished. For the investigation by spectroscopic ellipsometry, two similar samples were used except that they were thicker (0.5 mm) and roughened on one side to prevent disturbing interferences.

Optical transmittance (respective absorbance) are measured in the spectral range from 250 nm ($\sim$5 eV) to 500 nm ($\sim$2.5 eV) at temperatures from 5~K to 300~K. A double beam, double monochromator spectrophotometer (Perkin-Elmer Lambda 1050) is used in conjunction with a liquid-helium flow cryostat (Oxford OptistatCF) with UV-quartz windows. The spectral resolution is set to 2~nm. The linear polarization of the incident monochromatic light is chosen by a double polarizer insert equipped with Glan-Thomson polarizer crystals which are suitable for wavelengths above 240 nm. Attenuation of the reference beam (to 1 $\%$ transmission) is applied, allowing absorbance measurements up to 6. The absorption coefficient $\alpha$ is calculated from the measured transmittance \textit{T} using the expression~\cite{schroder+91wl}.
\begin{equation}
\alpha=-\frac{1}{d} \mathrm{ln}\left[\frac{\sqrt{(1-R)^{4}+4T^{2}R^{2}}-(1-R)^{2}}{2TR^{2}}\right],
\end{equation}
where \textit{d} is the thickness of the sample and \textit{R} the reflectance that we calculated by $R=(n-1)^{2}/(n+1)^{2}$ using the refractive index data obtained by spectroscopic ellipsometry measurements. For the determination of the onset of intrinsic optical absorption the square of the absorption coefficient was plotted versus the photon energy. Extrapolating the steep linear part of this plot to $\alpha^{2}=0$ yielded the photon energy we call in the following onset of optical absorption.

Spectroscopic ellipsometry was performed in the range from 1.5 eV to 6.5 eV at room temperature using a commercial ellipsometer (Horiba Jobin Yvon UVISEL). The complex reflectance ratio of the light polarized parallel and perpendicular to the plane of incidence ($r_{p}$ and $r_{s}$, respectively) is measured and can be expressed by $\frac{r_{p}}{r_{s}}=\mathrm{tan}\Psi e^{-i\Delta}$, where $\mathrm{tan}\Psi$ is the amplitude ratio and $\Delta$ the phase difference~\cite{azzam+87anh}. A large angle of incidence of 70$^{\circ}$ was used for the measurements, providing an electric field primarily aligned in the (100) or (001) planes of the wafers. The anisotropy of the dielectric function was determined by rotating the sample around its wafer normal to have one of the in-plane principal directions parallel to \textbf{E}. The evaluation of the spectroscopic ellipsometry data was performed under the assumption of a bulk semi-infinite homogeneous material thought to be appropriate for the surface-polished crystal wafers used here. This simple model allows direct transformation of the measured ellipsometric angles to the complex dielectric function and related parameters like the absorption coefficient by using the Fresnel equations. We do not include a correction to account for the non-ideal surface because of the usually necessary vague assumptions. Therefore, the absorption coefficients are overestimated in the range of weak absorption, but are reliable in the range of strong absorption. In this respect, the spectroscopic ellipsometry supplements the optical absorption spectroscopy that is only sensitive at the weak optical absorption onset. The onset of strong absorption was determined by evaluating plots of the squared absorption coefficient versus the photon energy.

\section{Electronic properties}
\label{electronic}
\begin{figure*} 
  \begin{center}
  \includegraphics[width=.95\textwidth]{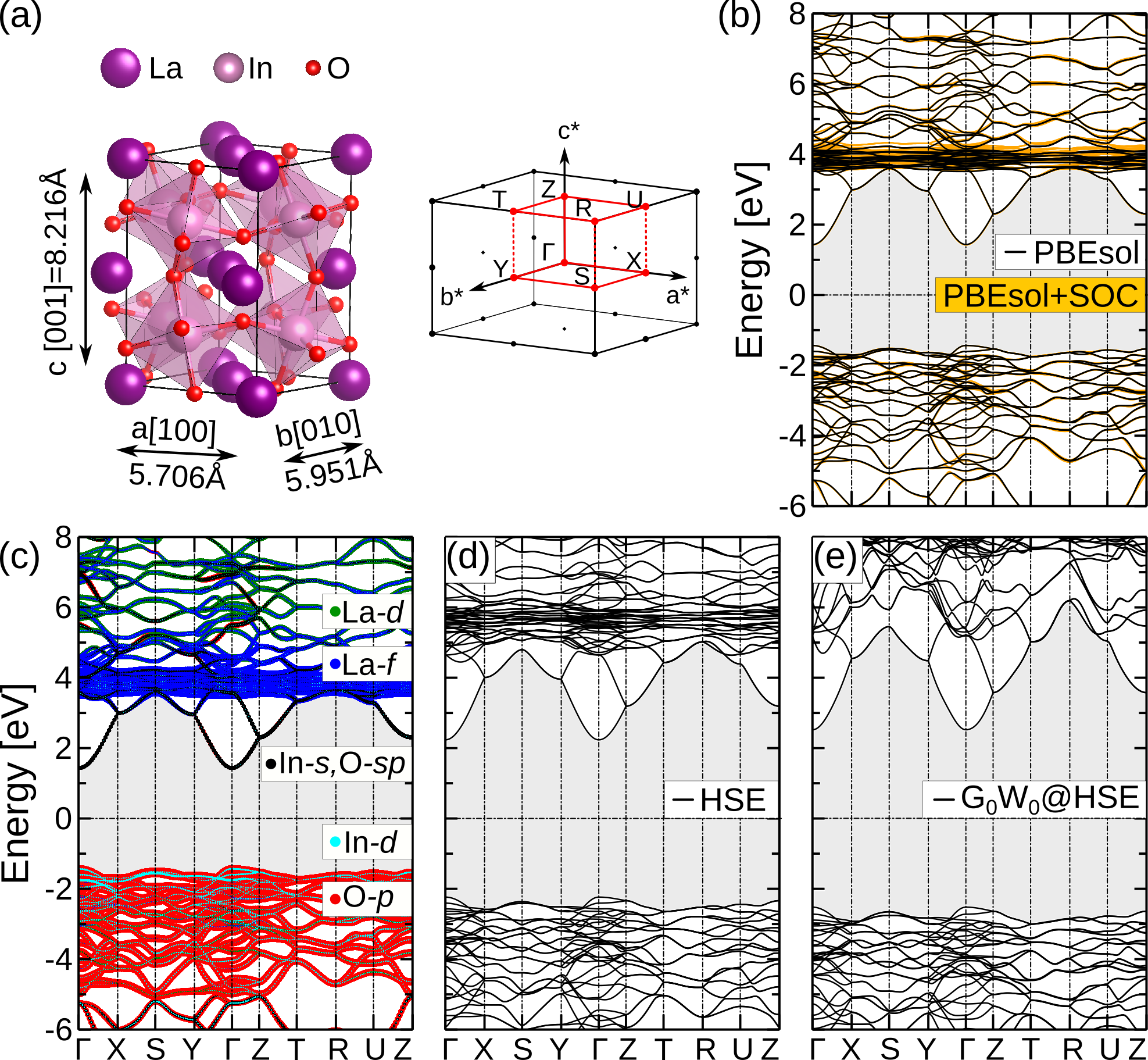}%
  \caption{(a) Primitive cell (left) of orthorhombic LaInO$_{3}$ and first Brillouin zone (right) with high-symmetry points and paths indicated in red; (b) PBEsol band structure with (orange) and without (black) spin-orbit coupling (SOC); (c) atomic characters projected onto the PBEsol band structure; (d) HSE06 and (e) $G_{0}W_{0}$@HSE06 band structures. In all panels, the mid gap is set to zero.}
  \label{fig:LIO-elec}
  \end{center}
\end{figure*}

LaInO$_{3}$ crystallizes in the orthorhombic space group Pbma, with four formula units per unit cell as shown in Fig.~\ref{fig:LIO-elec}(a)~\cite{park+03acsc,Zbigniew+20jcg,aytac+16pm}, and experimental lattice constants of \textit{a}= 5.722~\AA, \textit{b}= 5.938~\AA, and \textit{c}= 8.214~\AA~\cite{Zbigniew+20jcg}. Our calculated lattice constants of \textit{a}= 5.706~\AA, \textit{b}= 5.951 \AA, and \textit{c}= 8.216 \AA, obtained by PBEsol, are in a close agreement with the experimental values. Note that in literature, also an alternative notation is used, {\it i.e.}, space group Pnma with the longest axis \textit{b}, obtained by cyclical permutation of the axes~\cite{park+03acsc,Useong+15apl,Zbigniew+20jcg}.
 
In Fig.~\ref{fig:LIO-elec}, we present the band structure of LaInO$_{3}$ computed with PBEsol, HSE06, as well as $G_{0}W_{0}$@HSE06. The corresponding band gaps are reported in Table~\ref{tab:gaps}. According to PBEsol, both the valence band maximum (VBM) and the conduction band minimum (CBm) are located at the $\Gamma$ point, resulting in a direct band gap of 2.87 eV [see Fig.~\ref{fig:LIO-elec}(b)]. This value underestimates the experimental counterpart~\cite{Useong+15apl,jang+17jap,Zbigniew+20jcg}, by around 40\%. Since spin-orbit coupling has negligible effect on the electronic bands, as visualized in Fig.~\ref{fig:LIO-elec}(b), we omit it in the following calculations. The atomic character of the bands is highlighted by the color code in Fig.~\ref{fig:LIO-elec}(c). The valence bands above -5 eV are mainly dominated by O-\textit{p} orbitals, and In-\textit{d} states contribute to the highest VB around the $\Gamma$ and S points. The dispersive conduction band around the $\Gamma$ point is dominated by In-\textit{s} as well as O-\textit{s} and O-\textit{p} states (O-\textit{sp}). The La-\textit{d} orbital dominates the conduction bands above 4.5 eV, while the La-\textit{f} derived-bands appear at about 4 eV (2 eV above the CBm).

The HSE06 band structure is shown in Fig.~\ref{fig:LIO-elec}(d). Compared to PBEsol, the direct character of the gap at $\Gamma$ is preserved, its value of 4.45 eV is about 1.6 eV higher (see Table~\ref{tab:gaps}). The bands between the CBm and 3 eV above have almost the same shape in both cases. The main difference is in the location of the La-\textit{f}-derived bands that are shifted up to appear 1 eV higher in energy.

We also compute the quasiparticle band structure applying $G_{\mathrm{0}}W_{\mathrm{0}}$ on top of the HSE06 results. The latter method has turned out most reliable for computing the electronic properties of a wide range of oxides, including TCOs~\cite{Bechstedt+17jmr}. As shown in Fig.~\ref{fig:LIO-elec}(e), the QP correction increases the HSE06 band gap by 0.55 eV to about 5 eV. As electron-phonon (e-ph) effects are not considered at this level, the latter gap is sightly higher than the experimental values. We will get back to this point in Section~\ref{Sec:exp}~and~\ref{comparison}. Considering the whole QP band structure, the VB bands are almost the same as those given by HSE06, while the dispersion of the bands located between the 4 and 8 eV differs from both PBEsol and HSE06 results. Moreover, we find that the La-\textit{f}-derived bands are further shifted up to appear at 5 eV from the CBm, which is 2 eV more than in HSE06 [see Fig.~\ref{fig:LIO-elec}(d)]. This huge difference shows that a simple scissor shift of both PBEsol and HSE06 conduction bands will not be sufficient to study the optical properties of LaInO$_{3}$ in a high-energy range. However, in this work we are interested in the absorption-edge spectra that involve a small group of top valence and bottom conduction states. Comparing the bands computed with PBEsol and $G_{\mathrm{0}}W_{\mathrm{0}}@$HSE06, the states have the same ordering and similar dispersion in the vicinity of the Fermi level. This justifies the scissor approach for computing the optical spectra, as mentioned in the Section~\ref{method}. 

\begin{table}[h]
\centering
\caption{Electronic band gaps, $E_g$, and effective electron (hole) masses, m$_{e}^{*}$/m$^{0}$ (m$_{h}^{*}$/m$^{0}$), of orthorhombic LaInO$_{3}$ computed with PBEsol, HSE06, and $G_{0}W_{0}$@HSE06.}
\vspace{0.2cm}
 \begin{tabular}{c|c|c|ccccccc}
\hline
\hline
\multirow{2}*{LaInO$_{3}$} &                & \multicolumn{3}{c}{DFT} &  \multicolumn{1}{c}{$G_{0}W_{0}$@}&\\
\cline{3-4}
\cline{6-7}
                                    &      &PBEsol  &   HSE06          &  & HSE06 \\
\hline 
E$_{g}$ [eV]                         & $\Gamma$-$\Gamma$ &2.87 &   4.45          &  & 5.00 \\
 \cline{1-7} 
 \multirow{3}*{m$_{e}^{*}$/m$^{0}$}  &$\Gamma$-X& 0.53  &   0.46  &  &0.37 \\
 \cline{2-7}
 &$\Gamma$-Y&0.40   &   0.39  &  &0.31 \\
 \cline{2-7}
 &$\Gamma$-Z&0.50   &   0.48  &  &0.61 \\
 \cline{1-7} 
 \multirow{3}*{m$_{h}^{*}$/m$^{0}$}  &$\Gamma$-X&1.21   &   1.17   &  &1.05 \\
  \cline{2-7}
 &$\Gamma$-Y&0.59   &   0.52   &  &0.48 \\
  \cline{2-7}
 &$\Gamma$-Z&3.81   &   4.00   &  &3.46 \\
\hline
\hline
\end{tabular}
\label{tab:gaps}
\end{table}
In Table~\ref{tab:gaps}, we also report the values of the effective electron and hole masses. The smallest electron mass is found along the $\Gamma$-Y direction (corresponding to the lattice direction \textit{b}) with a value of 0.40 m$_{\mathrm{0}}$ from PBEsol and 0.39 m$_{\mathrm{0}}$ from HSE06. $G_{0}W_{0}$@HSE06 reduces this value to 0.31 m$_{\mathrm{0}}$ due to the changed dispersion around the CBm in the QP band structure. The values for the directions along $\Gamma$-Z and $\Gamma$-X are slightly higher, indicating anisotropic electron conductivity. The pronounced dispersion around the VBM along $\Gamma$-Y corresponds to a low effective hole mass of 0.59 and 0.52 m$_{\mathrm{0}}$ as obtained by PBEsol and HSE06, respectively. Employing $G_{0}W_{0}$@HSE06, it decreases to 0.48 m$_{\mathrm{0}}$. Overall, these values indicate that LaInO$_{3}$ can be a good candidate for both n- and p-type conductivity.
\begin{figure}
\begin{center}
\includegraphics[width=.49\textwidth]{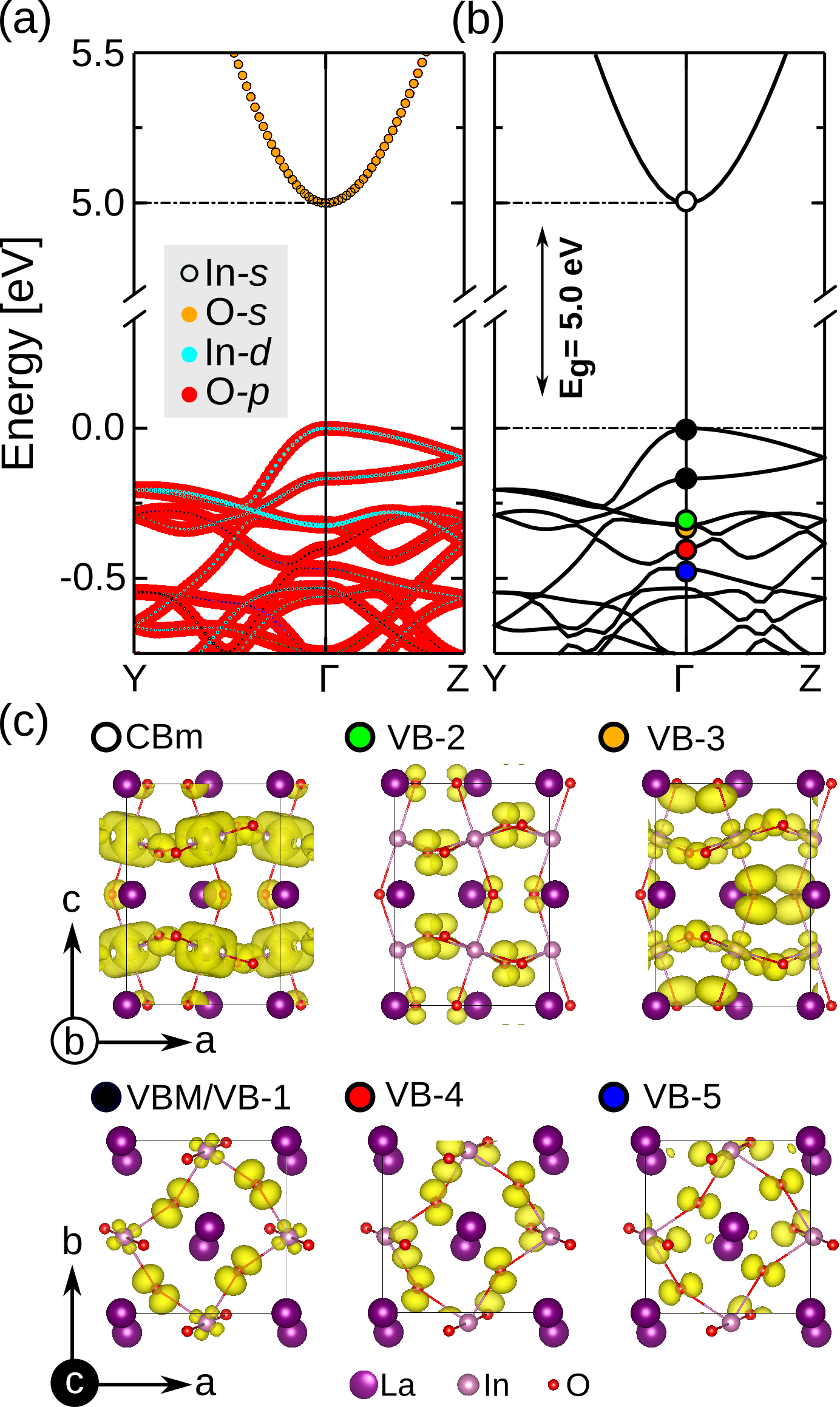}%
\caption{(a) Band character projected onto the PBEsol band structure along the Y-$\Gamma$-Z path in the BZ. The CB is shifted upwards by 2.13 eV to recover the $G_{0}W_{0}$@HSE06 band gap. (c) Square modulus of the Kohn-Sham wavefunction (yellow) of the VBM/VB-1, VB-2, VB-3, VB-4, VB-5, and CBm states at $\Gamma$, highlighted by black, green, orange, red, blue, and white circles, respectively, in the band structure shown in (b).}
\label{fig:LIO-ks}
\end{center}
\end{figure}

To analyze and understand the optical response of the material, more specifically the characteristic of the absorption onset, we depict in Fig.~\ref{fig:LIO-ks} the band structure in the vicinity of the band gap, highlighting the atomic characters [Fig.~\ref{fig:LIO-ks}(a)]. We also visualize the VB and CB wave functions at the $\Gamma$ point [Fig.~\ref{fig:LIO-ks}(b) and (c)]. As we can see, the valence bands are mainly formed by O-\textit{p} states. The VBM, VB-1, and VB-3 have additional contributions from In-\textit{d} orbitals. Such characteristic has also been reported for other TCOs such as In$_{2}$O$_{3}$ (bixbyite), Ga$_{2}$O$_{3}$ (bixbyite), and SnO$_{2}$ (rutile), where cation \textit{d}-states contribute to the top of the VB~\cite{sabino+15prb,sabino+17jpcm}. The bands with pure O-\textit{p} character are lower in energy and have odd parity. This is the case for VB-5, VB-4, and VB-2 that are formed by O-\textit{p} orbitals, exhibiting $p_{x}$ , $p_{y}$, and $p_{z}$ character, respectively. We note that VB-2 and VB-3 are energetically split at $\Gamma$. The CBm is built of In-\textit{s} and \textit{sp}-hybridized O states with even parity.

\section{Optical spectra}
\label{optical}
\subsection{BSE calculations}
\label{BSE}
\begin{figure}
 \begin{center}
\includegraphics[width=.47\textwidth]{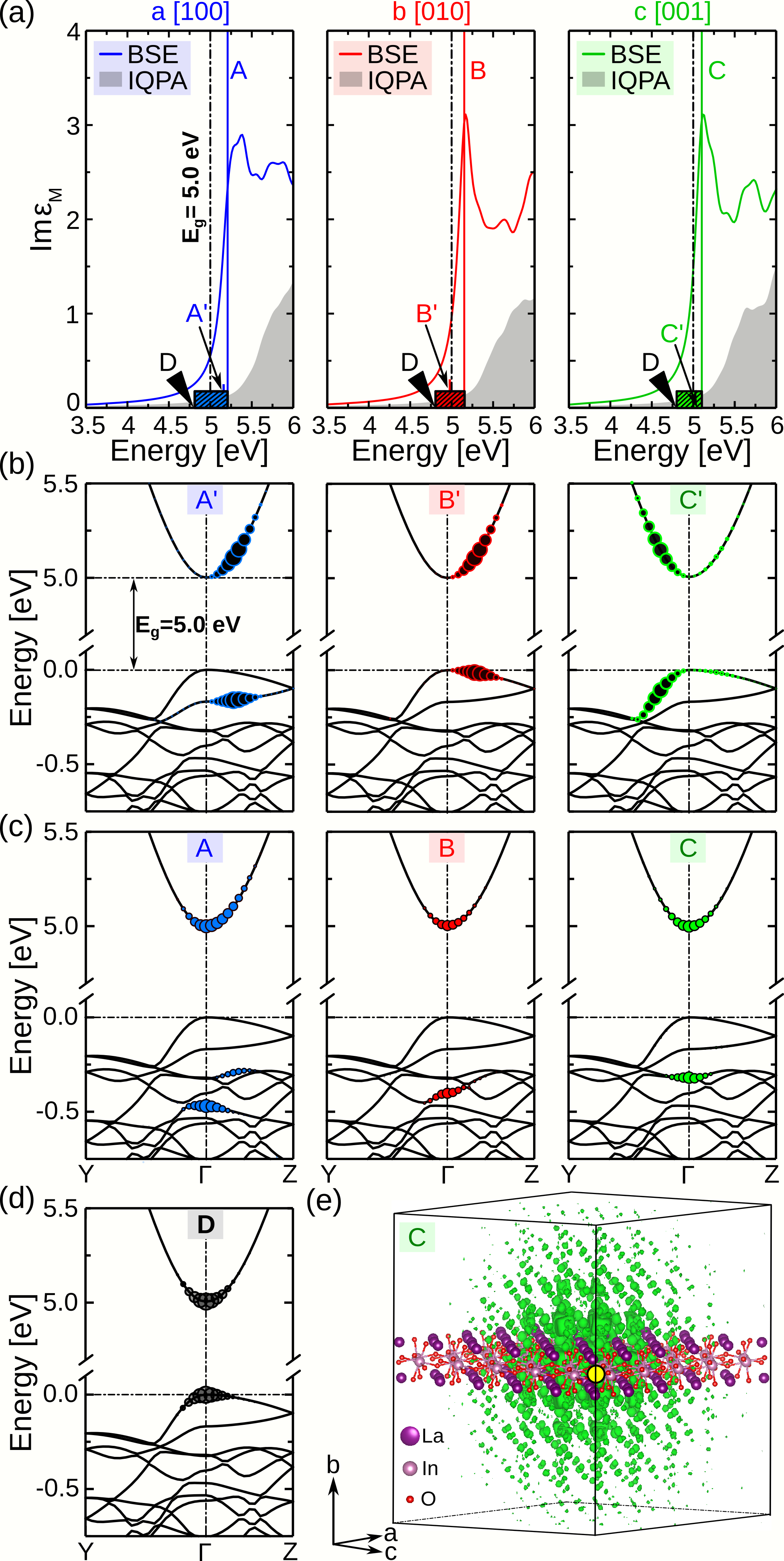}%
\caption{(a) Imaginary part of the macroscopic dielectric function of LaInO$_{3}$ for light polarization along [100] (blue), [010] (red), and [001] (green) direction as computed from BSE (solid line). The IQPA spectrum (shaded gray area) is shown for comparison. The position of the lowest-energy excitation D is marked by a black triangle, and the ranges of weak excitations are highlighted with dashed rectangles (BSE spectra). (b,d) Band contributions to the weak and (c) first intense excitations. (e) Exciton wavefunction of excitation C, showing the spatial electron distribution (green) with respect to hole being fixed at an oxygen atom (yellow spot). The crystal structure is shown locally to ease the visualization. }
\label{fig:LIO-optics}
 \end{center}
\end{figure}

In Fig.~\ref{fig:LIO-optics}(a), we plot the optical spectra of LaInO$_{3}$ for light polarization along [100], [010], and [001]. Focusing on the BSE results, we find the onset (marked as D) at about 4.80 eV {\it i.e.}, 0.2 eV below the fundamental gap, reflecting a sizeable electron-hole attraction. It stems from transitions from the top VB to the lowest CB around the $\Gamma$ point [see Fig.~\ref{fig:LIO-optics}(d)]. The reason for the very weak intensity is due to the fact that the direct transition at $\Gamma$ is dipole forbidden according to the parity of the VBM and CBm states. In Table~\ref{tab:parity}, we identify the type of transitions from the top five VBs to the CBm, according to their irreducible representation~\footnote{This analysis was carried out by using the $\mathrm{Irvsp}$ tool~\cite{irvsp} of the VASP code}. As the VBM and VB-1 have even parity, transitions to the CBm, that belongs to the A$_{1g}$ representation, are dipole forbidden. The same holds for the transition from the VB-3 to the CBm. Allowed transitions to the CBm come from the VB-2, VB-4, and VB-5, all exhibiting odd parity (B$_{1u}$, B$_{3u}$, and B$_{2u}$ representations, respectively). Thus, the optical onset is characterized by weak excitations from states around $\Gamma$ (along the $\Gamma$-X, $\Gamma$-Y, and $\Gamma$-Z paths), which are dipole allowed [see colored rectangles in Fig.~\ref{fig:LIO-optics}(a)]. Such transitions are found for light polarization along all directions, examples being excitation A' at 5.17 eV along [100], B' at 4.98 eV along [010], and C' at about 5.09 eV along [001] as indicated in Fig.~\ref{fig:LIO-optics}(a). Their origins in terms of initial and final states are highlighted in Fig.~\ref{fig:LIO-optics}(b). Note that these exciton weights (Eq.~\ref{eq:weight}) do not include the momentum matrix elements and thus do not provide information on the oscillator strength.

\begin{table}[h]
\centering
\caption{Irreducible representations of the top five valence bands at the $\Gamma$ point, determining whether transitions to the CBm are either dipole forbidden or allowed.}
\vspace{0.2cm}
 \begin{tabular}{|c|c|c|c|c|c|c||ccc}
\hline
Band &Symmetry at $\Gamma$& Transition to CBm \\
  \hline
  \hline
CBm & A$_{1g}$  &- \\
\hline
VBM & B$_{2g}$  &forbidden \\
VB-1 & A$_{1g}$ &forbidden \\
VB-2 & B$_{1u}$  &allowed \\
VB-3 & B$_{3g}$  &forbidden \\
VB-4 & B$_{3u}$  &allowed \\
VB-5 & B$_{2u}$  &allowed \\
\hline
\end{tabular}
\label{tab:parity}
\end{table}

Strong excitations only set in above the fundamental gap (vertical dashed line at 5.0 eV), {\it i.e.}, at about 5.22 (excitation A), 5.17 (excitation B), and 5.12 eV (excitation C) for light polarization along [100], [010], [001], respectively [see Fig.~\ref{fig:LIO-optics}(c)]. Their anisotropy arises from the fact that the transitions involve initial states from different bands, {\it i.e.}, VB-5, VB-4, and VB-2, respectively that are dominated by O-\textit{p} states of $p_{x}$ (blue), $p_{y}$ (red), and $p_{z}$ (green) character, respectively [see Fig.~\ref{fig:LIO-ks}(c)]. Only when the orientation of the O-\textit{p} orbitals coincides with the direction of light polarization, an intense excitation to the CB, mainly around the $\Gamma$ point [Fig.~\ref{fig:LIO-optics}(c)] appears. The binding energies of excitations A, B, and C are about 0.2 eV, as obtained by comparison with the corresponding IQPA transitions. To visualize the exciton wavefunction, we select excitation C as an example and display in Fig. \ref{fig:LIO-optics}(e) the electron distribution corresponding to a fixed position of the hole at an O site. The latter is delocalized with an extension larger than 20~\AA, mainly located on In-\textit{s} and O-\textit{s} orbitals.

We conclude by noting that this situation -- very weak excitations at the absorption onset and very strong ones above the fundamental band gap -- may lead to contradictory interpretations regarding the experimental determination of the optical gap, as we will see in Sections~\ref{Sec:exp} and~\ref{comparison}.
 
\subsection{Experiment} 
\label{Sec:exp}

\begin{figure}
 \begin{center}
\includegraphics[width=.49\textwidth]{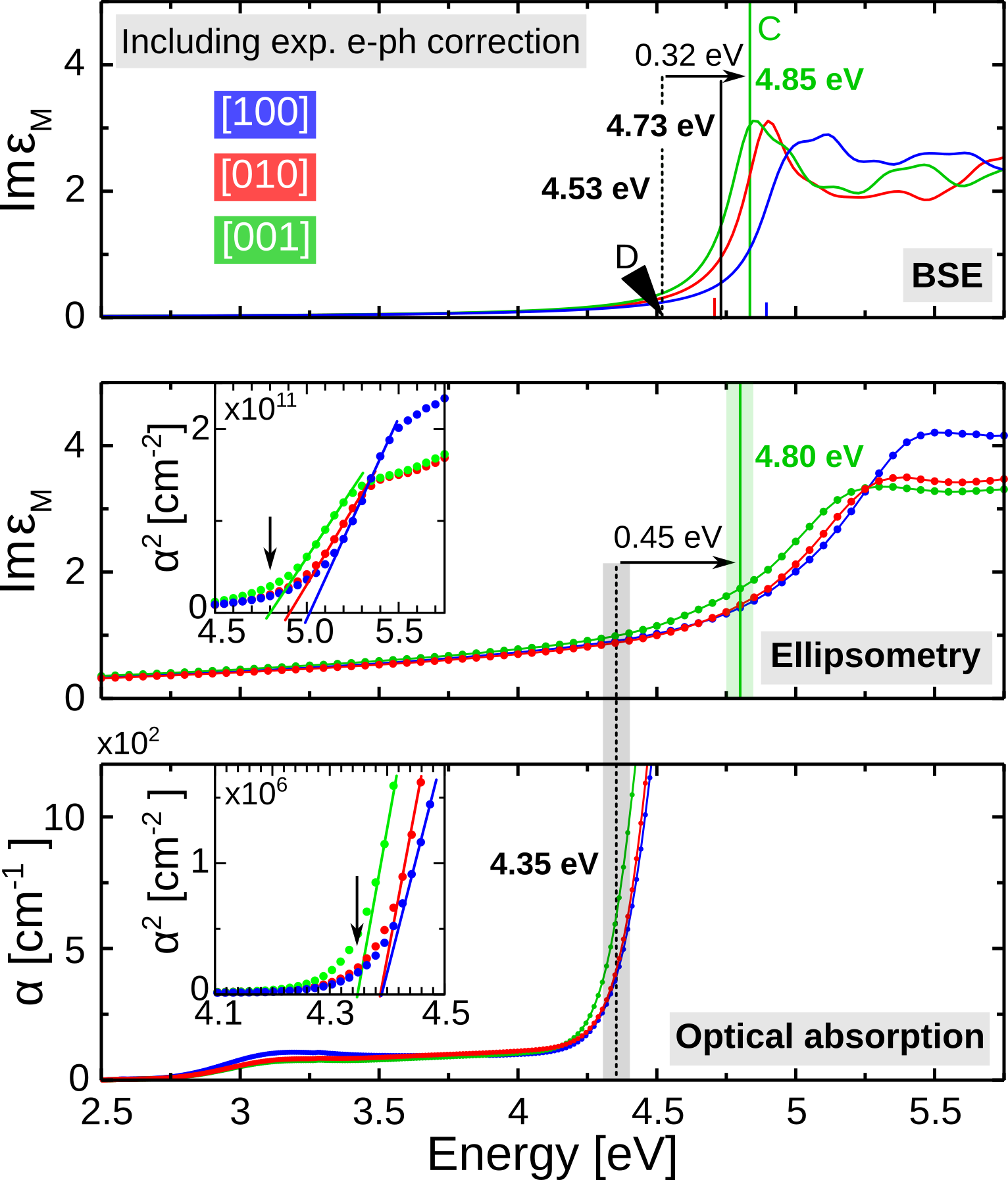}%
\caption{Top: Computed imaginary part of the macroscopic dielectric function of LaInO$_{3}$ (solid lines), for light polarization oriented along [100] (blue), [010] (red), and [001] (green) direction. The position of the lowest optical transitions (D) is marked with a black triangle. The first intense excitation ([001] polarization) is indicated by the vertical green line, the electronic QP gap by the vertical black line. Corrections due to ZPV and temperature effects are included such to correspond to 300 K. Absorption spectrum taken at room temperature of a single LaInO$_{3}$ crystal, obtained from ellipsometry (middle panel) and from optical absorption (bottom). The corresponding onsets are highlighted with green (solid) and black (dashed) vertical lines, respectively. The experimental error bar is estimated to be $\sim$0.1 eV, as indicated by the line thicknesses.}
\label{fig:LIO-exp}
 \end{center}
\end{figure}

In Fig.~\ref{fig:LIO-exp} (middle and bottom panels) we display the absorption spectra given by ellipsometry and optical absorption measurements, respectively. The BSE spectrum is shown in the top panel for comparison. At first glance, we recognize the same qualitative behavior concerning the anisotropy at the respective onset. However, strikingly, the absorption onset is not the same in the two experimental probes.
In ellipsometry, it appears at 4.80 eV ([001] direction) which is 0.45~eV above that given by optical absorption (4.35~eV). In Table~\ref{tab:exp-gaps}, we report these values. To shine light onto this discrepancy, we get back to the discussion of the BSE results.

\begin{table}[h]
\centering
\caption{Onsets of optical absorption [in eV] determined by spectroscopic ellipsometry at room temperature (300 K) as well as by absorption spectroscopy at 300 K and 5 K, for different polarization directions of the incident light.}
\vspace{0.2cm}
 \begin{tabular}{c|ccccccc}
\hline
\hline
\multirow{2}*{Polarization} &Ellipsometry&& \multicolumn{2}{c}{Absorption}\\
\cline{2-2}
\cline{4-5}
  &300 K &&300 K & 5 K \\
\hline
 \textbf{E} $\parallel$ [100] &5.0 && 4.39 & 4.52\\
\hline
 \textbf{E} $\parallel$ [010] &4.9 && 4.39 & 4.52\\
\hline
 \textbf{E} $\parallel$ [001] &4.8 && 4.35 & 4.46\\
\hline
\hline
\end{tabular}
\label{tab:exp-gaps}
\end{table}

\begin{figure}
 \begin{center}
\includegraphics[width=.49\textwidth]{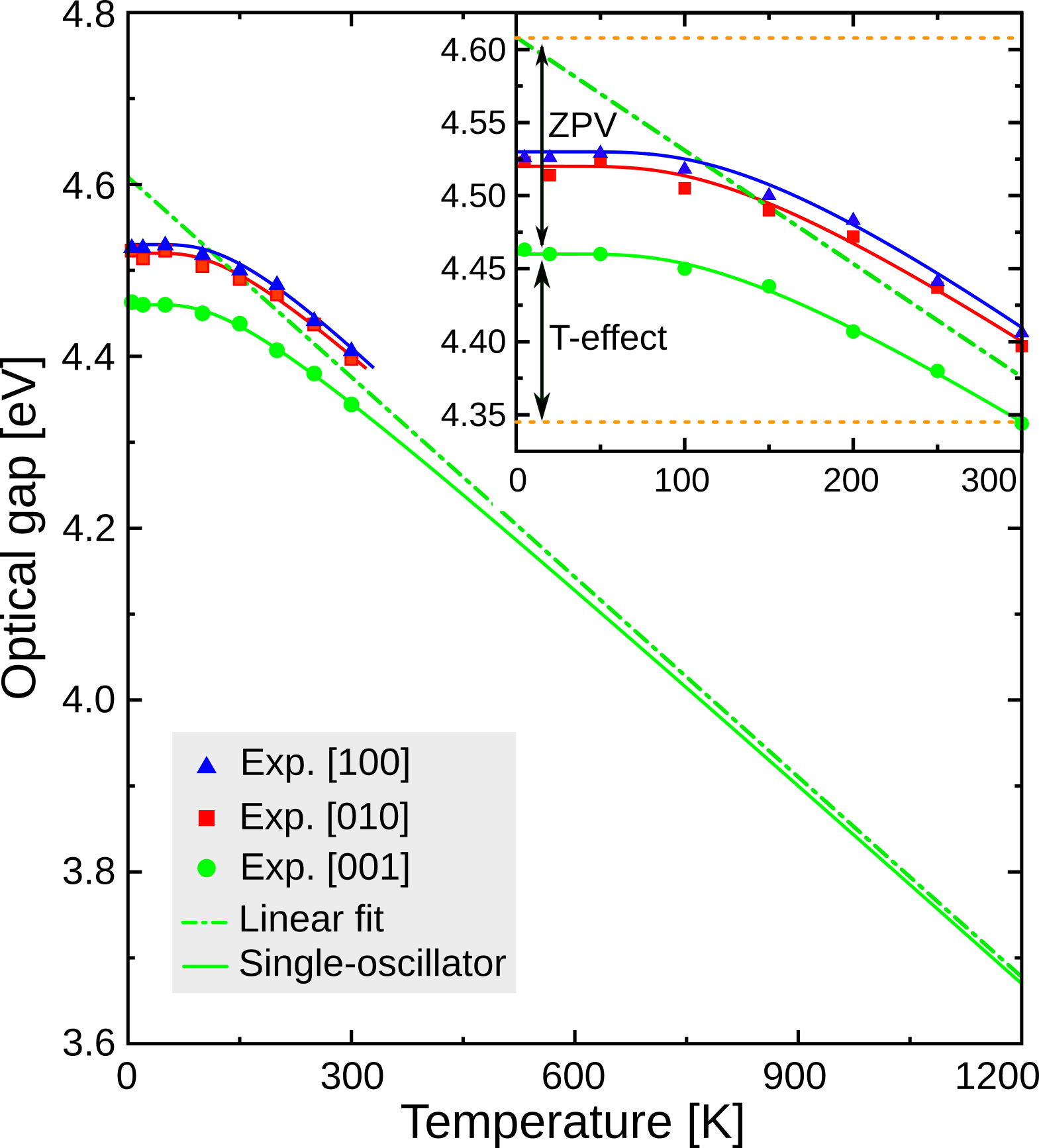}%
\caption{Temperature dependence of the optical gap obtained by optical absorption measurements. Blue, red, and green symbols represent the measured data for light polarization along [100], [010], and [001], respectively; the corresponding solid lines are the best fits based on a single-oscillator model. The dashed green line shows a linear fit for high temperatures. The inset zooms into the low-temperature region to distinguish between gap-renormalization due to temperature and ZPV.}
\label{fig:LIO-ZPV}
 \end{center}
\end{figure}

Before doing so, we first address vibrational effects, {\it i.e.}, renormalization of the band gap by zero-point motion and temperature. Not being considered in the calculated QP band gap, the absorption onset obtained from the BSE is to be expected higher than in experiment. We estimate these effects from optical absorption measurements performed below room temperature. The resulting temperature dependence of the optical gap is shown in Fig.~\ref{fig:LIO-ZPV}. 

The measured data are fitted using a single-oscillator model~\cite{Donnell+91apl}, expressed by
\begin{equation}
E_{\mathrm{g}}(T)=E_{\mathrm{g}}(0)-S\langle \hbar\omega \rangle \left[\mathrm{coth}\left( \frac{\langle \hbar\omega \rangle}{2k_BT}\right)-1\right],   
\end{equation}
where the parameter $E_{\mathrm{g}}(0)$ is the optical gap at zero temperature, $S$ a measure of the electron-phonon coupling strength, and $\langle\hbar\omega\rangle$ an average phonon energy \cite{Donnell+91apl,irmscher+14pssa}. Considering the whole temperature range, the best-fit parameters $E_{\mathrm{g}}(0)$=4.46~eV, $S$=4.5, and $\hbar\omega$=33~meV are obtained for polarization along the [001] direction. Comparable values are found for other directions. Extrapolating to 1200 K, the gap is by about 0.65 eV lower than at room temperature, reflecting the strong impact of electron-phonon coupling. We note that in other TCOs such as In$_{2}$O$_{3}$ and SrTiO$_{3}$, this effect is found to be even more pronounced, where the 
gap decreases by more than 1 eV~\cite{kok+15pssa,irmscher+14pssa}. The high-temperature part of the curve can be fitted with a linear regression, {\it i.e.}, $E_{\mathrm{g}}(0)-2Sk_BT+ S\langle \hbar\omega \rangle$~\cite{Donnell+91apl}. From this analysis, the zero-point vibration effects are estimated, amounting to $S\langle \hbar\omega \rangle\sim$0.15~eV. Thus, the optical gap (disregarding the electron-phonon contribution) is determined to be 4.62 eV at 0 K (see inset in Fig.~\ref{fig:LIO-ZPV}). Its decrease from 0 and 300 K is about 0.27 eV, where 0.15 eV is assigned to zero-point vibrations (ZPV) and 0.12 eV to temperature effects including lattice expansion. The latter value is comparable to those reported for other TCOs~\cite{irmscher+14pssa}. These values are included as corrections to the theoretical spectra for comparison with experiment (Fig.~\ref{fig:LIO-exp}).

\subsection{Comparison between theory and experiment}
\label{comparison}
In Table~\ref{tab:zpv-qp}, we report the QP gap, as given by $G_{0}W_{0}$@HSE06 at 0K, and the optical gap obtained by the BSE. The latter is defined by the lowest-energy excitation (D). In addition, we also provide the energy of the first intense excitation (E$^C$). These values need to be corrected in order to account for ZPV and temperature effects, both obtained from the analysis in Section \ref{Sec:exp}. At room temperature, the fundamental QP gap is 4.73 eV, while the optical gap is 4.53 eV, 0.32 eV below the first intense excitation C ([001] direction) at 4.85 eV. This difference, in fact, explains the seeming discrepancy between the two experimental probes (see Fig.~\ref{fig:LIO-exp}). More specific, ellipsometry, obviously only "sees" the high-intensity transitions, thus its absorption edge at $\approx$4.80 eV coincides with the first intense excitation (C, [001]) at 4.85 eV given by theory. In contrast, the optical absorption measurement is sensitive to weak excitations and thus its onset corresponds to the optical gap. Overall, there is excellent agreement with our calculations, well within the experimental and theoretical error bars (see Fig.~\ref{fig:LIO-exp}). With these findings, we can rule out the origin of the weak excitations to originate from phonon-assisted (indirect) transitions~\cite{sabino+17jpcm,irmscher+14pssa,weiher+66jap}. Nevertheless, the absorption onset may be enhanced by the presence of point-defects, that are evident from the absorption at about 2.8 eV which lower the crystal symmetry.

\renewcommand{\arraystretch}{1.3}
\begin{table}[h]
\centering
\caption{QP gap, E$_g^{QP}$, as given by $G_{0}W_{0}$@HSE06 at 0K and optical gap, E$_g^{opt}$, as obtained by the BSE [in eV]. In addition to the lowest-energy excitation (D), defining the optical gap, also the energy of the first intense excitation (E$^C$) is provided. The corresponding values including corrections to account for ZPV and temperature effects are depicted as well.}
\vspace{0.2cm}
 \begin{tabular}{c|cc|cc|cc}
Theory &             E$_g^{QP}$  & E$_g^{QP-corr}$   & E$_g^{opt}$ &    E$_g^{opt-corr}$  &E$^C$ &  E$^{C-corr}$\\
\hline
0 K                 & 5.0   &  4.85 & 4.80  & 4.65 &  5.12&  4.97\\
300 K               &     & 4.73 &&  4.53 &&     4.85  \\
\end{tabular}
\label{tab:zpv-qp}
\end{table}
  
 Finally, we deduce the fundamental gap from the optical absorption. By adding the excitonic binding energy of 0.2 eV obtained by BSE to the experimental optical absorption onset, we arrive at $E_{g}=$4.55 eV at room temperature.

\section*{Summary and Conclusions}
We have presented a detailed study of the electronic and optical properties of the orthorhombic perovskite LaInO$_{3}$, from both first-principles many-body calculations and experiment. Employing the self-energy correction on the HSE06 results ($G_{0}W_{0}$@HSE06), we have found a direct electronic QP gap of about 5 eV. The effective electron (hole) mass is estimated to be about 0.31 (0.48) m$_{\mathrm{0}}$ along the $\Gamma$-Y direction. These low values suggest the material useful for both n- and p-type conductivity. Spin-orbit coupling is shown to have a negligible effect on the electronic structure in the vicinity of the Fermi energy. Temperature-dependent optical absorption measurements show the optical gap to decrease by about 0.12 eV going from 5 to 300 K. Below that, zero-point vibration effects are estimated to amount to a further reduction of 0.15 eV. From the solution of the BSE, we have found that the absorption onset is characterized by weak excitations, stemming from the vicinity of the band edges, but the transition exactly at the VBM ($\Gamma$ point) being dipole forbidden. The electron-hole binding energy of the lowest exciton is about 0.2 eV. Our findings are fully inline with optical absorption measurements that capture also excitations with low oscillator strength. Intense excitations only set in by about 0.32 eV higher in energy. Including corrections due to ZPV and temperature effects, the energy of the first intense excitation obtained theoretically coincides with the absorption edge obtained by ellipsometry at 4.8 eV which is unable to provide information on the optical gap. All three methods find excellent agreement concerning the anisotropy of the optical spectra, most strikingly, the BSE and ellipsometry results, finding response to light polarized along [001] lowest and along [100] highest in energy. 

Input and output files can be downloaded free of charge from the NOMAD Repository~\cite{drax-sche19jpm} at the following link: \url{https://dx.doi.org/10.17172/NOMAD/2021.01.10-1}.

\subsection*{Acknowledgment}  
This work was supported by the project BaStet (Leibniz Senatsausschuss Wettbewerb, No. K74/2017) that is embedded in the framework of GraFOx, a Leibniz ScienceCampus, partially funded by the Leibniz Association. We acknowledge the North-German Supercomputing Alliance (HLRN, project bep00078) for providing HPC resources. We thank Albert Kwasniewski (IKZ) for the x-ray crystallographic orientation of the LaInO$_{3}$ samples.

%
\end{document}